# Role of focusing distance in picosecond laser-induced Cu plasma spectra

Linyu Chen,[1] Hu Deng,[1] Zhixiang Wu,[1] Zhonggang Xiong,[1,2] Jin Guo,[1,3] Quancheng Liu,[1] Akwasi Danso Samuel,[1] Liping Shang,[1,*]

[1]School of Information Engineering, Southwest University of Science and Technology, Mianyang 621010, China.
[2]School of Mechanical Engineering, Guilin University of Aerospace Technology, Guilin 541004, China.
[3]Tianfu College of Southwestern University of Finance and Economics, Mianyang 621000, China.
*shangliping@swust.edu.cn

**Abstract:** To study the effects of focusing distance on the characteristics of copper plasma, a picosecond laser was utilized to ablate a pure copper plate to generate a plasma spectrum. Following numerous experiments on the subject, three significant factors have been determined: lens focal length, pulse energy and the lens-to-sample distance. These factors were employed to analyze the spectral intensity, plasma temperature and electron density in the local thermodynamic equilibrium (LTE) and optically thin condition. Due to the shielding effects of mixed plasma, the strongest spectral intensity can be obtained in the pre-focused case rather than on the focus, no matter how much beam irradiance was employed. The more intensive the beam irradiance is, the more the optimal position is distant from the focal point. Similarly, the evolution of plasma temperature and electron density was shown a peak in the pre-focused case, which is consistent with the trend of spectral intensity. For the case of extremely high irradiance (on the focus), the shielding effects become more apparent and the resultant above three factors decreased sharply. When a longer-focal-length lens was employed, the spectral intensity exhibited an obvious bimodal trend. In the pre-focused case, a longer-focal-length lens is helpful to eliminate the effects of the roughness of the target surface compared with a shorter one. Finally, the assumed LTE was validated by McWhirter relation, plasma relaxation time and diffusion length, and the optically thin condition also validated by spectral intensity ratio. We hope this work could be an important reference for the future design of highly optimized experiments for Calibration-Free Laser-Induced Breakdown Spectroscopy (CF-LIBS).

## 1. Introduction

Alongside the advancements in high-performance lasers and novel chemometrics methods in the last 30 years, laser-induced breakdown spectroscopy (LIBS) has become an attractive technique, with much potential, for the analysis of elements of materials [1]. LIBS offers distinct advantages for in-situ, real-time, standoff and contactless detection, and it requires less sample preparation and consumption [2]. The technique is widely applied to the fields of energetic materials [3], metallurgical engineering [4], space exploration [2], industry manufacture [5], food safety [6], environmental monitoring [7], agricultural product [1], resources and energy [8], archaeology [9], and clinical diagnostics [10]. However, the stability and repeatability of the spectral intensity, along with the detection accuracy, are easily altered by the wavelength of the input light, pulse duration and its energy, the distance of lens-to-sample and the target characteristics.

In an attempt to enhance the spectral intensity, and thus improve the analytical precision, a larger number of insightful studies have been offered within the past ten years. For example, Fahad *et al*. used a pulse of wavelength 1064 nm and 532 nm to analyze the plasma characteristics of marble calcite [11]. When a femtosecond laser is used to ablate a copper target, the spectral intensity of Cu (I) and Zn (I) is higher when the laser is circularly polarized rather than linearly polarized [12]. The dependence of the spectral intensity of Si on the position of a lens, at different sample temperatures, is investigated by Zhang *et al* [13]. While Guo *et al*. analysed the influence of the distance between the lens and the sample on spectral intensity, and the vibrational temperature of the CN band, by using a nanosecond pulse to ablate polymethyl methacrylate [14]. Harilal *et al* investigated the spot size effects on propagation dynamics [15] and conversion efficiency [16] of laser-produced plasma in vacuum and argon atmosphere, the results show that the sharpening of plasma plume depends heavily on the spot size. An improvement of up to three times has been achieved by Carvalho *et al*. in the measurement precision, when the particle size of plant pellets is less than 20 μm and larger than 150 μm [17]. By combining spatial confinement and double-pulse, the intensity of the Cr lines has a prominent enhancement factor of 168.6 [18]. Of course, there are other methods for enhancing spectral intensity, such as magnetic fields [19], sample heating [20], and nanoparticle coating [21].

The above work mainly uses a nanosecond or a femtosecond laser to generate a plasma spectrum. Compared with these two lasers, picosecond laser has appropriate pulse width with higher pulse energy, which can increase the ablation volume and spectral intensity (for femtosecond pulse), and may alleviate the effect of plasma shielding as well (for nanosecond pulse). It's well known that the pulse duration and beam irradiance of a laser are two crucial parameters in the process of laser-matter interaction. The longer the duration, the more heat effects on plasma or the surface of a target. When a femtosecond pulse is used to illuminate the target, the phenomenon of target heating and vaporing will occur a few picoseconds after the laser pulse, and the target plasma-laser interaction never comes [22]. As for nanosecond LIBS, the enhanced spectrum is attributed to plasma reheating and increased ablation mass because of longer duration or enhanced illuminate energy, and the effect of plasma shielding is a considerable factor in the procedure of plasma formation. Compared with femtosecond, the pulse energy of a picosecond laser is up to a few hundred mJ, which can increase the spectral intensity caused by plasma reheating or ablation mass [23]. The mass removal from the target is more reproductive, and the intensity-to-background ratio of a spectral line is higher when using a picosecond rather than a nanosecond laser [24].

Generally, plasma shielding is an undesirable phenomenon in LIBS application (especially in nanosecond LIBS). For femtosecond and nanosecond laser-induced plasma, the variation of line intensity and plasma temperature is shown a similar trend with the ranged lens-to-sample distance, while the plasma formation and shielding effects are not completely identical. Due to huge differences in pulse duration, the features of picosecond pulse or plasma may different from the femtosecond/nanosecond pulse. This work attempts to analyze the influence of focusing distance on the characteristics of picosecond laser-induced Cu plasma. A series of experiments are conducted in this work using a set of fixed focal lengths, pulse energies and lens-to-sample distances. This is achieved by using three lenses of focal length, $f$ = 50 mm, 100 mm, and 150 mm, while the intensity and its optimal focusing distance are thoroughly investigated for pulse energy ranges from 12 mJ to 44 mJ. Under the local thermodynamic equilibrium (LTE) and optically thin condition, we have observed the evolution of plasma temperature and electron density with increased focusing distance.

## 2. Materials and Methods

A schematic for picosecond laser-induced plasma spectroscopy is shown in Fig. 1. The copper plasma is produced with a Q-switched Nd: YAG laser (SL234, Ekspla) operating at FWHM (Full Width at Half Maximum) duration of 120 ps. This picosecond laser can output at a maximum energy of 250 mJ with a stability of 1.5%, when working at its fundamental wavelength of 1064 nm. The laser pulse is reflected by the mirror (denoted by M1 in Fig. 1), and then it is focused on the surface of a copper plate via a lens, L1. This lens is fixed on a vertical linear translation 1D-stage (NRT150, Thorlabs), while the copper plate is located on a 3D-stage. On the right-hand side of Fig. 1, the parameter H represents the lens-to-sample distance; this can range from 40 mm to 160 mm by a movement of the 1D-stage. To ensure the accurate H, the size of ablated spot produced by low pulse energy (~0.15mJ) on a piece of photographic paper is measured by an industrial microscope (LV100D, Nikon) and used to estimate the practical focal position of the lens with $f$ = 50 mm. The Cu plasma spectrum is collected via two plano-convex lenses, L2 and L3 (with focal lengths 150 mm and 100 mm, respectively), and then coupled to a multichannel fiber spectrometer (AvaSpec-ULS2048-8-USB2, Avantes). The wavelength of this spectrometer ranges from 200 nm to 1070 nm, and the resolution is greater than 0.06 nm for 312 nm - 564 nm. The trigger and delay of the picosecond laser and spectrometer are controlled manually by the digital delay/pulse generator DG645, with a burst mode to generate an external 8 Hz trigger.

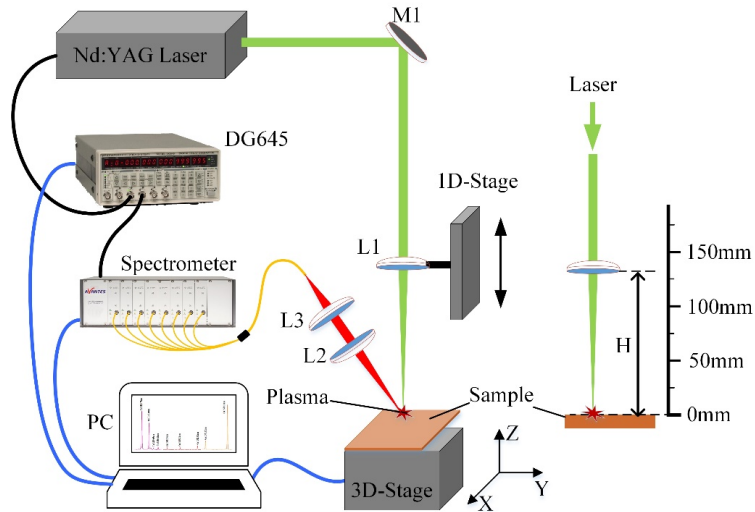

Fig.1. Schematic for picosecond laser-induced plasma spectroscopy with a variable lens-to-sample distance.

Before the experiment is begun, a calibration light source (AvaLight-CAL-Mini, Avantes) is used to calibrate the sensitivity of the spectrometer. According to the method reported in the literature [25], the instrumental broadening (~0.045 nm) is estimated by a pair of Hg lines (576.96 and 579.06 nm) emitted from this lamp. The pure copper plate is moved at a constant speed to ensure that each shot provides a fresh point. Each spectrum data is an average of 10 continuous shots. The interval between the laser pulse and the actual trigger of the spectrometer is set to 1.3 μs, and the gate width of the spectrometer exhibits a minimum value of 1.05 ms. The whole experiment was carried out in air atmosphere, and air humidity ranges from 28% to 32% at room temperature.

# 3. Results and discussion

## 3.1 Spectral Intensity at A Fixed Focal Length

At first the pulse energy is set to 20 mJ, and a lens with $f$ = 150 mm is chosen to illuminate the surface of copper plate. When the focusing distance (H) is 149 mm (pre-focused case), the Cu plasma spectrum is given in Fig. 2. In this circumstance, there are nine spectral lines with reasonably strong intensities. By referring to the NIST database and related report [26], those spectral lines clearly all relate to the Cu atomic lines at wavelengths of 324.75, 327.40, 329.05, 330.80, 510.55, 515.32, 521.82, 570.02 and 578.21 nm.

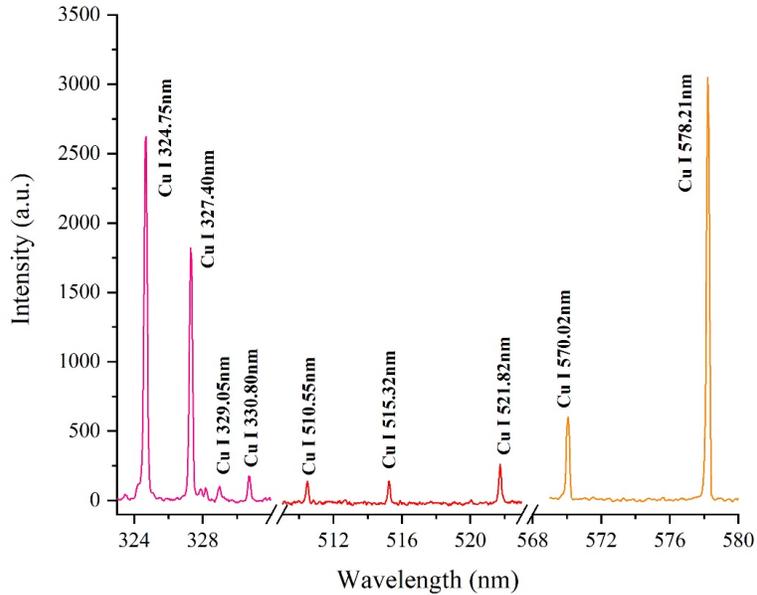

Fig. 2. Graph showing the plasma spectra for a Cu plate at H = 149 mm ($f$ = 150 mm, pulse energy is 20 mJ).

When H ranged from 144.6 mm to 156 mm, the spectra intensities of the three strongest lines, relating to Cu I at 324.75, 327.40 and 578.21 nm, are shown in Fig. 3. We can determine from this graph that, in comparison to the continuous spectral intensities of the background, these three spectral lines are much larger and easily distinguishable from the rest. It transpires that two intensity peaks arise in the pre-focused (H = 149 mm) and defocused (H = 151 mm) cases, but the pre-focused one has significantly higher intensity compared to the defocused one. For the pre-focused case (H < 150 mm), as the H increased, the coupling efficiency between the laser pulse and the sample is increased firstly due to the reduced spot size, resulting in a stronger spectrum and an opaque plasma with higher density. Once the plasma density reaches a critical value, a thin layer forming above the target surface will absorb the subsequent energy and prevent the laser from propagating to the target, which is the phenomenon of plasma shielding [26]. Referring to the related report on laser-material interaction [27], the pulse duration of ~120 ps is sufficient to maintain the target plasma encounters the rest incident energy. When the focus point is located on the target surface (H = 150 mm, i.e. on the focus), the plasma shielding has dominated the energy and the spectral intensity drops sharply. From this observation, we can conclude that the ablated copper surface is modified by both the beam irradiance of the pulse and the effects of plasma shielding.

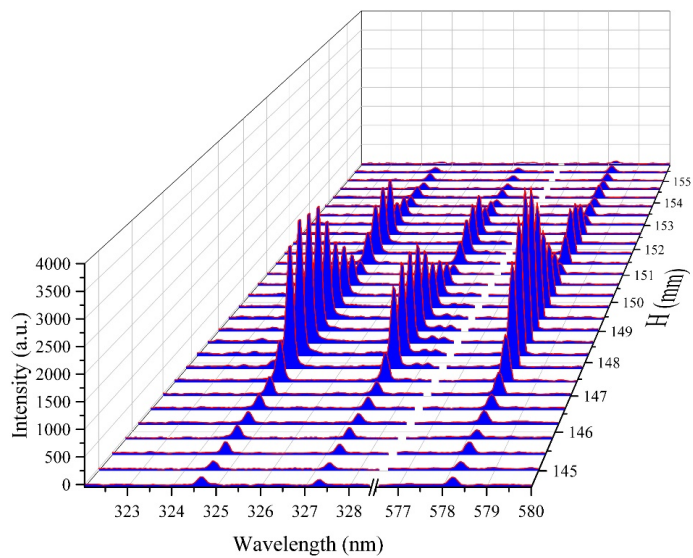

Fig. 3. Graph showing the spectra of the three Cu(I) lines with H ($f$ = 150 mm, pulse energy is 20 mJ).

## 3.2 Spectral Intensity at Different Focal Lengths

To observe the optimal lens-to-sample distance (with high line intensity) and plasma shielding at the different focal lengths, the pulse energy is again set to 20 mJ but the lens L1 now has either $f$ = 50 mm or 100 mm. The atomic line intensities of Cu at 324.75, 327.40, 510.55, 515.32, 521.82, 570.02 and 578.21 nm change with H as shown in Fig. 4.

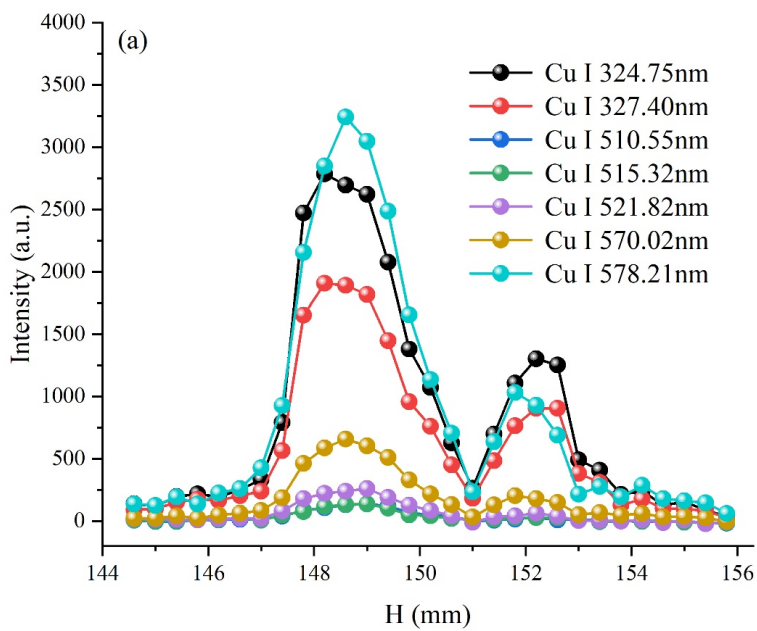

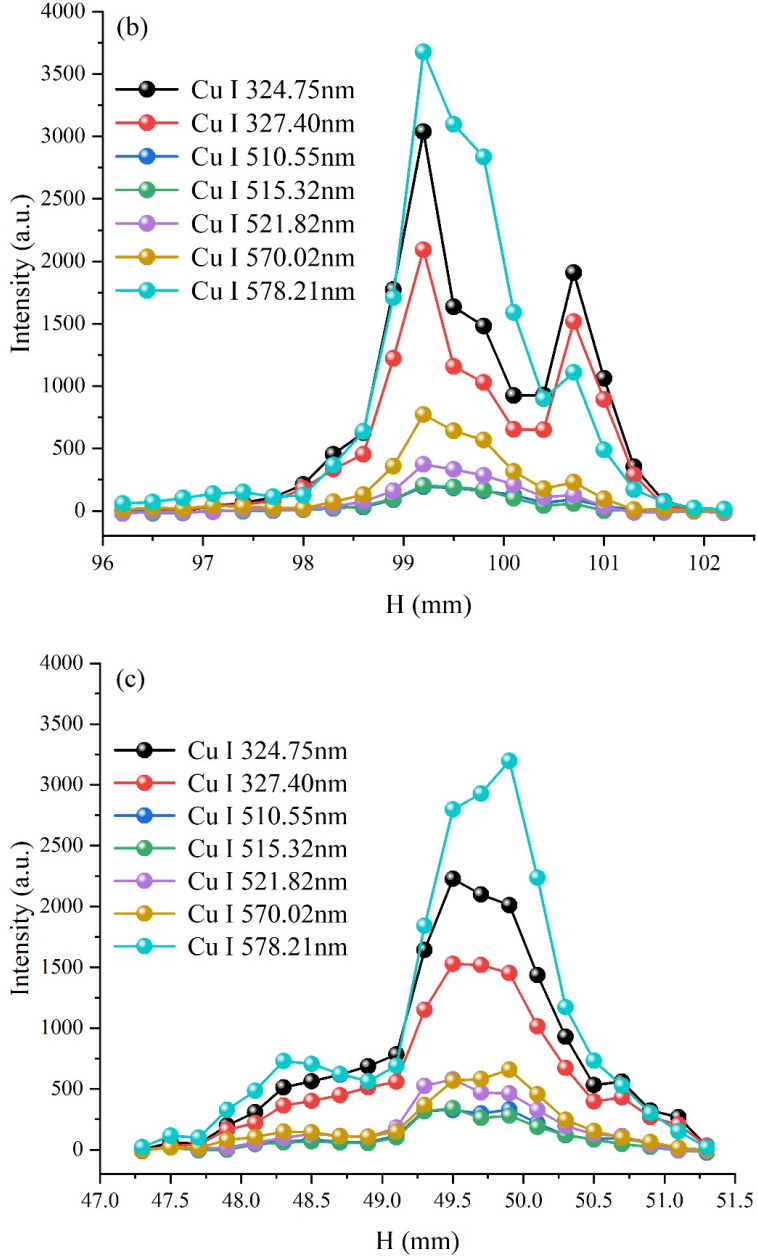

Fig .4. Graphs showing that the intensities of Cu atomic lines (i.e. Cu(I) at 324.75 nm, 327.40 nm, 510.55 nm, 515.32 nm, 521.82 nm, 570.02 nm and 578.21 nm) change with *H*, when using the lens L1 with a focus length of (a) 150 mm, (b) 100 mm, and (c) 50 mm. The pulse energy is 20 mJ.

Examining the system using a lens with $f$ = 150 mm, when H changes from 144.5 mm to 149 mm (at the pre-focused case), the intensity of the seven spectral lines of Fig. 4(a) initially increase more and more rapidly. This is because the increasing H reduces the spot size on the target surface, causing a non-linear increase in the beam irradiance. In the case of low beam irradiance, as long as the electron density is above the critical threshold, an increased beam irradiance will lead to more energy absorption in the region of laser-supported detonation wave (LSDW) [28], increasing the ablated volume of the copper surface and plasma emission. The maximum spectral intensity occurs at H = 148.6 mm, which is about 2.3 times the intensity of the focal point (H = 150 mm). While using the lens with $f$ = 100 mm, the highest intensity can be obtained at H = 99.2 mm, and its spectral intensity is about 2.6 times that of the focal point.

When H is within reach of 150 (±1) mm (approximately on the focus), and the sample gradually moves from within the focal point to outside it, the intensity of the spectral lines show a similar sharp downward trend, which reaches the minimum value at H = 151 mm. This means that, when the copper plate is near to the focal point, the laser pulse has an extremely high irradiance. The plasma plume above the copper surface becomes opaque to the laser pulse [29], which will produce a strong shielding effect and results in almost no energy transferring to the surface of copper. It greatly reduces the ablation mass on the copper surface, and ultimately leads to a weak spectral intensity.

For the defocus case (typically H > 150 mm), another peak intensity point is reached at H = 152 mm. This is because the decreased beam irradiance produces a low electron density, which strengthens the coupling efficiency between the mixed plasma and the target, resulting in the above three parameters increasing again [28]. In this work, the high-fluence laser transmits through the air atmosphere, the self-focusing effect might occur and help to relieve the defocusing effect of the lens and enhance the irradiate fluence on the target surface. When H > 152 mm, the increased size of the focusing spot leads to a declined beam irradiance, which gradually reduces the ablation mass and the spectral intensity until the beam irradiance is below the breakdown threshold of copper.

Following the examination of the three lenses with different focal lengths, we determined that when f = 100 mm or 150 mm, the spectral intensities show an obvious bimodal trend with increasing H, while this trend does not occur when f = 50 mm. The width of the first peak indicated that when the lens is pre-focused, a longer-focal-length lens is helpful to eliminate the effects of the roughness of the target surface on line intensity compared with a shorter-focal-length lens [30]. Meanwhile, the distinct plasma shielding is easily observed mainly due to the air breakdown above the target surface. Besides, the ranged spot size on the target surface is another essential factor for inducing plasma shielding. In the air atmosphere, the plasma plume structure is long and narrow when using a long-focal lens but relatively flat when a short length is employed. The geometric size of the plasma plume impacts the coupling efficiency between the laser pulse and the target surface [16], which directly alters the ablation mass and spectral intensity.

### 3.3 Plasma Emission Dependence on Pulse Energy

It is known that the beam irradiance directly affects the ablation mass of the target surface, which in turn alters the coupling efficiency between the target and the incident beam, and finally induce the changed spectral intensity of the plasma. In this section, the intensity change of line 578.21 nm is selected as the one used to analyze in-depth the effects of the pulse energy. Taking the lens with a focal length of 150 mm as an example, with the pulse energy ranging from 12 mJ to 44 mJ, the intensity of the spectral line at 578.21 nm with different H values is shown in Fig. 5. It's observed that the spectral intensity is enhanced with the increased pulse energy, and the maximum spectral intensity becomes increasingly stronger. The maximum spectral intensity and its corresponding H values indicate that the best location of the focusing lens is the pre-focused case, rather than on the focal point with the highest energy fluence. It can be established that the higher the pulse energy, the farther the target (with the maximum spectral intensity) is from the focal point.

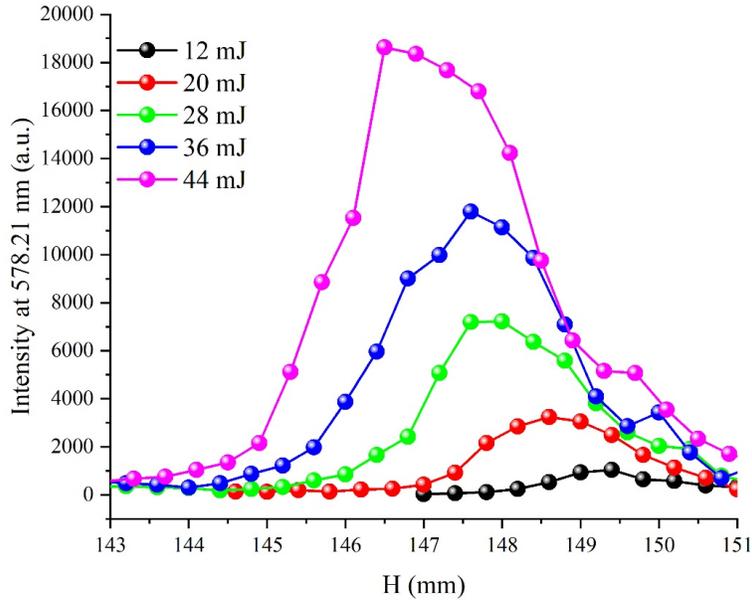

Fig .5. Evolution of spectral intensity with H at different pulse energies when $f$ = 150 mm.

With pulse energy ranging from 12 mJ to 44 mJ, the corresponding maximum intensity of 578.21 nm are shown in Fig. 6. After optimizing the lens-to-sample distance, no matter which of the three lenses is used, the pulse energy increases linearly and the spectral intensity also shows a linear increasing trend. When the pulse energy is more than 20 mJ, and a lens with $f$ = 100 mm is used, the optimal spectral intensity is more sensitive to a change in the pulse energy.

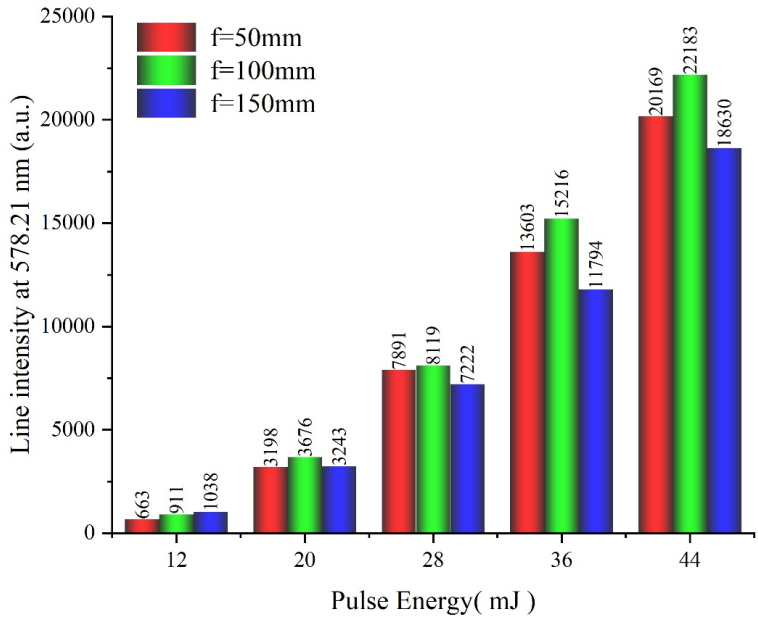

Fig .6. Maximum spectral intensity of Cu(I) at 578.21 nm for various pulse energies.

With increasing pulse energy, the optimal lens-to-sample distances (with the maximum spectral intensity) are calculated and presented in Fig. 7. We found that, when the same lens is used to both focus and ablate the copper surface, the larger the pulse energy the closer the lens is to the target with maximum spectral intensity. Taking the lens with a focal length of 150 mm as an example, when the pulse energy is 12, 20, 28, 36 and 44 mJ, comparing the positions of the focal point, the lens moves downwards by 0.6, 1.4, 2, 2.4 and 3.5, respectively. Under the same pulse energy, the maximum spectral intensity is obtained when the distance of the target surface to the focus is greater (if a longer focal length is used). Taking pulse energy of 36 mJ as an example, when lenses of $f = 50$ mm, $f = 100$ mm, and $f = 150$ mm are used, relative to each focal point, the copper plate moves forward by 0.6, 1.5 and 2.4 mm, respectively.

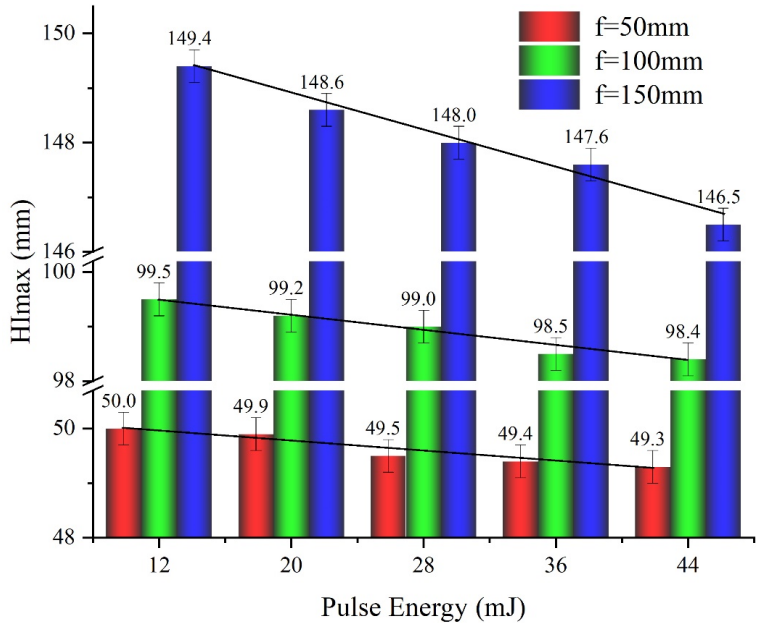

Fig .7. Graph showing the lens-to-sample distance when the maximum spectral intensity varies with pulse energy (HImax represents the lens-to-sample distance with maximum spectral intensity. The error bar is 0.3 mm, coming from the procedure of H calibration.).

Once the pulse energy and lens have been determined, it is well known that the farther the sample is from the focal position, the larger the spot size and the lower the beam irradiance, with the highest beam irradiance at the focal point. However, the spectral intensity at the focal point is not the strongest (as shown in Fig. 4) and is much lower than the pre-focused position. Therefore, we can conclude that the higher the beam irradiance the more serious the effects of plasma shielding. With a large increase in pulse energy, it is supposed that the increased rate of plasma shielding is lower than the rate of energy, though the corresponding effects of plasma shielding are enhanced. Finally, the higher laser energy irradiates to the copper surface, producing a larger ablated volume and emitting an intense spectrum. On the other hand, with the increased pulse energy, more energy will propagate through the plasma and ablate the surface of a target, with more ablated volume and enhanced spectrum. While the copper position of maximum spectral intensity will be closer to the lens (as shown in Fig. 6) and the size of the ablated spot will become larger. Besides, the absorbed energy will heat the plasma, which contributes to the intensity of the emission spectrum.

In the course of plasma shielding, the absorption factor of energy in plasma is tightly related to the combined effects of the ablation spot size, beam irradiance, plasma geometry, and plasma shielding [29]. Any severe shielding weakens the coupling efficiency between the laser pulse and the target surface and reduces the ablation mass on the target surface, resulting in a weaker emission spectrum intensity.

### 3.4 Plasma Temperature and Plasma Density

The existence of plasma shielding will induce the plasma to absorb the rest of pulse energy. Besides the emitted spectrum, the other characteristics, such as the plasma temperature and electron density, will be affected by the changed ablation energy or reheat energy. In practical applications of LIBS, the local thermodynamic equilibrium (LTE) and optically thin are generally assumed to evaluate the plasma characteristics. When the above assumptions are established, the self-absorption effects in the procedure of plasma emission can be ignored [31]. The plasma temperature $T$ (K) is determined from the fitted slope of the Boltzmann plot,

$$\ln \frac{I_1 \lambda_1}{A_1 g_1} - \ln \frac{I_2 \lambda_2}{A_2 g_2} = -\frac{1}{K_b T}(E_1 - E_2) \tag{1}$$

where $\lambda_1$ and $\lambda_2$ represent the central wavelength of two atomic spectral lines with their corresponding spectral intensities being $I_1$ and $I_2$, $A_1$ and $A_2$ represent the spontaneous transition probabilities of spectral lines $\lambda_1$ and $\lambda_2$ from high energy levels to low energy levels, $g_1$ and $g_2$ are the degeneracy of the upper levels of $E_1$ and $E_2$, respectively, and $K_b$ is the Boltzmann constant (eV/K).

Three Cu atomic lines at 510.55, 515.32 and 521.82 nm are used to determine the precise plasma temperature. The above lines not only match the following four physical features, such as a non-resonance line with less self-absorption, higher transition probability, individual line without interference and deformation, and enough difference of the upper-level energy of two lines, but also within the same channel with identical coupling efficiency between the plasma and the optical fiber. The spectroscopic parameters of the above three Cu atomic lines can be obtained in the NIST database.

Generally, Stark broadening of the spectral linewidth is mainly caused by the collisions of electrons and ions. Both the linear and the quadratic Stark effect are encountered in spectroscopy. However, only the hydrogen atom and H-like ions exhibit the linear Stark effect, whereas all other atoms exhibit the quadratic Stark effect. When calculating the electron density, the FWHM of $H_\alpha$ line profile can be calculated easily with greater accuracy than Cu atomic lines. Stark broadening was assumed to fit a Lorentz line shape. The electron density $N_e$, a function of FWHM $\Delta\lambda$, is determined by the following expression [32],

$$N_e = 8.02 \times 10^{12} \left(\frac{\Delta\lambda}{\alpha}\right)^{\frac{3}{3}} \text{cm}^{-3} \tag{2}$$

where $\lambda$ is the FWHM of the line in Å, $\alpha$ is the reduced Stark profiles in Å as tabulated by Kepple (i.e., $\alpha$ is 0.0149, line 2 in Table 1, page 322.) [33].

Here, taking the lens with a focal length of 150 mm as an example. When pulse energy is 44 mJ, the curves of plasma temperature and electron density at the corresponding H values are shown in Fig. 8. It is found that the peak of evolution is located around H = 146.5 mm, which is corresponding to the first peak of spectral intensity in Fig .5.

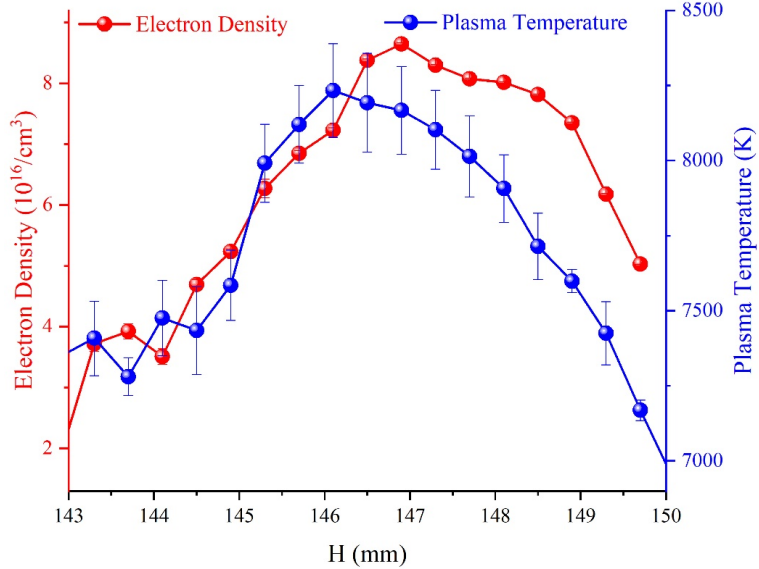

Fig .8 Evolution of electron density and plasma temperature at different H.

When H ranged from 143 mm to 146.5 mm, the copper surface absorbed the pulse energy and induce the increased temperature of the electrons and lattices. If the electrons absorb enough incident photons, it will emit along the opposite direction of the incident beam with high speed and high density, and break down the free air and produce the initial weak air plasma (weak spectral line Hα was observed) above the target surface [27]. Once the beam irradiance or electron density reaches a certain critical threshold, the shielding effects will occur in the mixed plasma consisting of target plasma and air plasma. Due to the shielding effects, the mixed plasma will absorb a part of the pulse energy through the electron-ion inverse-bremsstrahlung or the electron–neutral inverse-bremsstrahlung processes[31]. The spectral intensity, plasma temperature and electron density began to show a peak value, which means that the plasma shielding is relatively weak. When H > 146.5 mm, the enhanced energy fluence will aggravate the shielding effects. It implies that less and less energy can reach the target surface, and that mixed plasma will absorb more energy to strengthen the procedure of ionization and collision between the different particles. Although the increased energy is converted into the internal thermal energy of the mixed plasma through inverse-bremsstrahlung radiation, the plasma temperature has shown a declined slope. We deem that the narrow geometry of mixed plasma aroused by the smaller size of ablation spot alleviate the reheating effects on plasma. When the target is located inside the focal point, the curve of changing plasma temperature suggested that the rise in plasma temperature is governed by the ablation mass, rather than the plasma reheat.

### 3.5 Validity of Assumption

LTE refers to the existence of thermal equilibrium within a small region of the plasma. In LTE plasma, the collisional processes of excitation and de-excitation dominate the radiative processes, which require the electron density to be following the McWhirter relation [34],

$$N_e \geq 1.6 \times 10^{12} T^{1/2} (\Delta E)^3 \tag{3}$$

where $\Delta E$ represents the difference between the upper and lower energy (eV) levels of the spectral line. According to the physical parameters mentioned in the last section, the minimum electron density can be approximately calculated as $0.202*10^{16}$/cm$^3$ when the plasma temperature is up to 8233 $K$. It is lower than the experimentally obtained values at any H in Fig. 8, the McWhirter criterion is satisfied.

It is assumed that the plasma is homogeneous and stationary in the McWhirter relation. While the McWhirter relation is just a necessary condition for the dominance of LTE plasma, rather than a sufficient condition. Given the plasma size in this actual experiment, it needs to consider the transient properties, i.e. plasma relaxation time and diffusion length, in the homogeneous and inhomogeneous plasma, respectively. For relaxation time, it means that the evolution rate of homogenous and transient plasma must be slower than the electron collisions and can relax the plasma back to equilibrium. This criterion is given by the following inequality [34],

$$\frac{T(t+\tau_{rel}) - T(t)}{T(t)} \ll 1 \tag{4}$$

$$\frac{N_e(t+\tau_{rel}) - N_e(t)}{N_e(t)} \ll 1 \tag{5}$$

where the $\tau_{rel}$ is the plasma relaxation time, which can be expressed as [34],

$$t_{rel} = \frac{6.3 \times 10^4}{N_e f_{mn} \langle g \rangle} \Delta E_{nm} (K_b T)^{\frac{1}{2}} \exp\left(\frac{\Delta E_{nm}}{K_b T}\right) \tag{6}$$

where $f_{mn}$ is the oscillator strength of the m–n transition (ranging from 0.42 to 0.54, Table 1.1 in literature [35]). $\langle g \rangle$ is the effective Gaunt factor (dimensionless, ~ 04, Fig. 1 in literature [36]). $\Delta E_{nm}$ is the energy difference between the upper and lower level. For line 521.82 nm, the relaxation time $\tau_{rel}$ is estimated at ~ 3.3 nanoseconds, which is much shorter than the observation time of the spectrometer in this work. It means that the LTE is satisfied when the plasma is homogenous and transient in this work.

To consider the valid LTE as overall as possible, the third criterion, diffusion length, should be estimated in the inhomogeneous and transient plasmas. It requires that the diffusion length ($\Delta x$) of species, during a period of the order of the relaxation time ($\tau_{rel}$) to the equilibrium, is shorter than the variation length of temperature and electron number density in the plasma. The condition for LTE at the position $x$ can be expressed as [34],

$$\frac{T(x+\Delta x) - T(x)}{T(x)} \ll 1 \tag{7}$$

$$\frac{N_e(x+\Delta x)-N_e(x)}{N_e(x)} \ll 1 \qquad (8)$$

the diffusion length $\Delta x$ is expressed as [34],

$$\Delta x \approx \frac{1.4\times 10^{12}}{N_e}(kT_e)^{\frac{3}{4}}\left(\frac{\Delta E_{nm}}{M_A f_{mn}\langle g\rangle}\right)^{\frac{1}{2}}\exp\left(\frac{\Delta E_{nm}}{2kT_e}\right) \qquad (9)$$

where the $M_A$ is the relative mass of the species considered. Generally, the diffusion length must be at least one order of magnitude shorter than the plasma dimension. For Cu element, the value of $M_A$ is 29. The diffusion length $\Delta x$ is estimated as ~1.92 μm, while the plasma diameter can be estimated at ~1 mm. It means that the LTE is validated when the plasma is inhomogeneous and transient in this work.

As assumed in the calculating procedure of plasma temperature and electron density, the spectral lines are free from the self-absorption in optically thin plasma. For a plasma spectrum under the optically thin condition, the intensity relationship between two atomic lines must submit Equation (1). Especially for the lines originating from the same upper level in a specific element with the identical ionization stage, the theoretical doublet intensity ratio can be expressed by simplification of Equation (1) as follows [37],

$$\frac{I_1}{I_2}=\frac{A_1 g_1 \lambda_2}{A_2 g_2 \lambda_1} \qquad (10)$$

where the three factors are the atomic inherent parameters of the transitions, which is independent of the experimental parameters.

Among the three Cu atomic lines used to calculate the plasma temperature, the double lines 515.32 nm and 521.82 nm have the same upper level. We can calculate the experimental intensity ratio of 521.82 nm to 515.32 nm($I_{521.82\,nm}/I_{515.32\,nm}$) at different H. The experimental value of $I_{521.82\,nm}/I_{515.32\,nm}$ ranges from 1.68 to 1.81, which is close to the theoretical value of 1.85 obtained from Equation (6). Therefore, the observed plasma can exist in an optically thin condition. In conclusion, the assumption of LTE and optically thin condition is validated by minimum electron density, slower evolution and theoretically spectral intensity ratio.

## 4. Conclusion

This work examines the spectral intensity characteristics of copper plasma under a LIBS technique using various conditions, such as different lens focal lengths, pulse energies, and focusing distances. The above phenomenon shows that, with an increasing lens-to-copper surface distance, the difference of the focal length and pulse energy leads to the beam irradiance of the ablating spot changing under various behaviours. The beam irradiance and size of the ablated spot not only changes the geometric structure of the plasma, but also modifies the particle density inside the plasma. The combined effects of these critical factors result in a change of the coupling efficiency between the laser pulse and sample surface, which also affects the spectral intensity, plasma temperature and electron density. When the pulse duration exceeds 5 ps, the pulse will be interacting with plasma and result in plasma reheating. The source of plasma may be different, a shorter picosecond laser will produce air plasma above the surface of a target, while the nanosecond laser will generally produce vapour-plasma from the removal mass [23]. Under the effects of plasma shielding, the variation trend of spectral intensity with the focusing distance is consistent with the results of 8 ns and 50 fs duration in literature [29] and [38], respectively. In this work, we deem that the main reason for the hump in Fig .4 is similar to that of nanosecond pulse, namely the plasma reheating between the laser and mixed plasma, rather than the effects of self-focusing, defocusing and refocusing in the procure air ionization by a femtosecond laser. As for the second peak outside the focal point, the self-focusing effects will help to enhance the coupling efficiency between the laser pulse and the sample. Finally, three necessary and sufficient factors, the McWhirter relation, plasma relaxation time and diffusion length are used to validate the assumed LTE. And the theoretically spectral intensity ratio was employed to validate the optically thin condition. The present results provide a further understanding of the influence of beam irradiance on spectral intensity in picosecond LIBS. We hoped that this study could be a simple and effective method to improve detecting sensitivity without extra optical elements in the next work.

**Data Availability.** According to the management agreement of our laboratory, the data used to support the findings of this study are available from the corresponding author upon reasonable request.

**Conflicts of Interest.** The authors declare no conflicts of interest.

**Funding Statement.** This work was supported by the National Natural Science Foundation of China (Grant No. 11872058); the Sichuan Science and Technology Program of China (Grant No. 2019110, and 2019YFG0114); the Guangxi Natural Science Foundation of China (Grant No. 2019GXNSFBA185013); the National Defense Basic Scientific Research Program of China (Grant No. JCKY2018404C007, JSZL2017404A001, and JSZL2018204C002).

**References**

[1] J. Peng, Y. He, J. Jiang et al., "High-accuracy and fast determination of chromium content in rice leaves based on collinear dual-pulse laser-induced breakdown spectroscopy and chemometric methods," *Food chemistry*, vol. 295, pp. 327–333, 2019.
[2] N. L. Lanza, S. M. Clegg, R. C. Wiens et al., "Examining natural rock varnish and weathering rinds with laser-induced breakdown spectroscopy for application to ChemCam on Mars," *Applied Optics*, vol. 51, no. 7, B74-82, 2012.
[3] I. Gaona, J. Serrano, J. Moros et al., "Range-adaptive standoff recognition of explosive fingerprints on solid surfaces using a supervised learning method and laser-induced breakdown spectroscopy," *Analytical chemistry*, vol. 86, no. 10, pp. 5045–5052, 2014.
[4] C. Pan, X. Du, N. An et al., "Quantitative Analysis of Carbon Steel with Multi-Line Internal Standard Calibration Method Using Laser-Induced Breakdown Spectroscopy," *Applied Spectroscopy*, vol. 70, no. 4, pp. 702–708, 2016.
[5] N. Zhao, D. Lei, X. Li et al., "Experimental investigation of laser-induced breakdown spectroscopy assisted with laser-induced fluorescence for trace aluminum detection in steatite ceramics," *Applied Optics*, vol. 58, no. 8, pp. 1895–1899, 2019.
[6] B. A. Alfarraj, H. K. Sanghapi, C. R. Bhatt et al., "Qualitative Analysis of Dairy and Powder Milk Using Laser-Induced Breakdown Spectroscopy (LIBS)," *Applied Spectroscopy*, vol. 72, no. 1, pp. 89–101, 2018.
[7] Y. Qu, Q. Zhang, W. Yin et al., "Real-time in situ detection of the local air pollution with laser-induced breakdown spectroscopy," *Optics Express*, vol. 27, no. 12, A790-A799, 2019.
[8] M. Burger, P. J. Skrodzki, L. A. Finney et al., "Remote Detection of Uranium Using Self-Focusing Intense Femtosecond Laser Pulses," *Remote Sensing*, vol. 12, no. 8, p. 1281, 2020.
[9] R. Gaudiuso, M. Dell'Aglio, O. de Pascale et al., "Laser-induced breakdown spectroscopy of archaeological findings with calibration-free inverse method: Comparison with classical laser-induced breakdown spectroscopy and conventional techniques," *Analytica chimica acta*, vol. 813, pp. 15–24, 2014.
[10] Q. Wang, G. Teng, X. Qiao et al., "Importance evaluation of spectral lines in Laser-induced breakdown spectroscopy for classification of pathogenic bacteria," *Biomedical optics express*, vol. 9, no. 11, pp. 5837–5850, 2018.
[11] M. Fahad and M. Abrar, "Laser-induced breakdown spectroscopic studies of calcite ($CaCO_3$) marble using the fundamental (1064 nm) and second (532 nm) harmonic of a Nd: YAG laser," *Laser Physics*, vol. 28, no. 8, p. 85701, 2018.
[12] Q. Wang, A. Chen, W. Xu et al., "Signal improvement using circular polarization for focused femtosecond laser-induced breakdown spectroscopy," *Journal of Analytical Atomic Spectrometry*, vol. 34, no. 6, pp. 1242–1246, 2019.
[13] D. Zhang, A. Chen, Q. Wang et al., "Effect of lens focusing distance on laser-induced silicon plasmas at different sample temperatures," *Plasma Science and Technology*, vol. 21, no. 3, p. 34009, 2019.
[14] K. Guo, A. Chen, and X. Gao, "Influence of distance between target surface and focal point on CN emission of nanosecond laser-induced PMMA plasma in air," *Optik*, vol. 208, p. 164067, 2020.
[15] S. S. Harilal, "Influence of spot size on propagation dynamics of laser-produced tin plasma," *Journal of Applied Physics*, vol. 102, no. 12, p. 123306, 2007.
[16] S. S. Harilal, R. W. Coons, P. Hough et al., "Influence of spot size on extreme ultraviolet efficiency of laser-produced Sn plasmas," *Applied Physics Letters*, vol. 95, no. 22, p. 221501, 2009.

[17] G. G. A. de Carvalho, D. Santos, L. C. Nunes et al., "Effects of laser focusing and fluence on the analysis of pellets of plant materials by laser-induced breakdown spectroscopy," *Spectrochimica Acta Part B: Atomic Spectroscopy*, 74-75, pp. 162–168, 2012.

[18] L. B. Guo, B. Y. Zhang, X. N. He et al., "Optimally enhanced optical emission in laser-induced breakdown spectroscopy by combining spatial confinement and dual-pulse irradiation," *Optics Express*, vol. 20, no. 2, pp. 1436–1443, 2012.

[19] P. Liu, R. Hai, D. Wu et al., "The Enhanced Effect of Optical Emission from Laser Induced Breakdown Spectroscopy of an Al-Li Alloy in the Presence of Magnetic Field Confinement," *Plasma Science and Technology*, vol. 17, no. 8, pp. 687–692, 2015.

[20] Y. Wang, A. Chen, Q. Wang et al., "Influence of distance between focusing lens and target surface on laser-induced Cu plasma temperature," *Physics of Plasmas*, vol. 25, no. 3, p. 33302, 2018.

[21] X. Wen, Q. Lin, G. Niu et al., "Emission enhancement of laser-induced breakdown spectroscopy for aqueous sample analysis based on Au nanoparticles and solid-phase substrate," *Applied Optics*, vol. 55, no. 24, p. 6706, 2016.

[22] A. Valenzuela, C. Munson, A. Porwitzky et al., "Comparison between geometrically focused pulses versus filaments in femtosecond laser ablation of steel and titanium alloys," *Applied Physics B*, vol. 116, no. 2, pp. 485–491, 2014.

[23] J. P. Singh and S. N. Thakur, *Laser-Induced Breakdown Spectroscopy*, Elsevier, Amsterdam, London, 2007.

[24] K. L. Eland, D. N. Stratis, T. Lai et al., "Some Comparisons of LIBS Measurements Using Nanosecond and Picosecond Laser Pulses," *Applied Spectroscopy*, vol. 55, no. 3, pp. 279–285, 2001.

[25] L. Yue-Hua, C. Ming, L. Xiang-Dong et al., "The mechanism of effect of lens-to-sample distance on laser-induced plasma," *Acta Physica Sinica*, vol. 62, no. 2, p. 25203, 2013.

[26] R. Junjuri, S. A. Rashkovskiy, and M. K. Gundawar, "Dependence of radiation decay constant of laser produced copper plasma on focal position," *Physics of Plasmas*, vol. 26, no. 12, p. 122107, 2019.

[27] S. S. Mao, X. Mao, R. Greif et al., "Dynamics of an air breakdown plasma on a solid surface during picosecond laser ablation," *Applied Physics Letters*, vol. 76, no. 1, pp. 31–33, 2000.

[28] J. A. Aguilera, C. Aragón, and F. Peñalba, "Plasma shielding effect in laser ablation of metallic samples and its influence on LIBS analysis," *Applied Surface Science*, 127-129, pp. 309–314, 1998.

[29] J. Wang, X. Li, H. Li et al., "Lens-to-sample distance effect on the quantitative analysis of steel by laser-induced breakdown spectroscopy," *Journal of Physics D: Applied Physics*, vol. 53, no. 25, p. 255203, 2020.

[30] X. Li, W. Wei, J. Wu et al., "The Influence of spot size on the expansion dynamics of nanosecond-laser-produced copper plasmas in atmosphere," *Journal of Applied Physics*, vol. 113, no. 24, p. 243304, 2013.

[31] O. A. Ranjbar, Z. Lin, and A. N. Volkov, "Effect of the spot size on ionization and degree of plasma shielding in plumes induced by irradiation of a copper target by multiple short laser pulses," *Applied Physics A*, vol. 126, no. 5, p. 95, 2020.

[32] A. M. El Sherbini, H. Hegazy, and T.M. El Sherbini, "Measurement of electron density utilizing the Hα-line from laser produced plasma in air," *Spectrochimica Acta Part B: Atomic Spectroscopy*, vol. 61, no. 5, pp. 532–539, 2006.

[33] P. Kepple and H. R. Griem, "Improved Stark Profile Calculations for the Hydrogen Lines Hα, Hβ, Hγ, and Hδ," *Physical Review*, vol. 173, no. 1, pp. 317–325, 1968.

[34] G. Cristoforetti, A. de Giacomo, M. Dell'Aglio et al., "Local Thermodynamic Equilibrium in Laser-Induced Breakdown Spectroscopy: Beyond the McWhirter criterion," *Spectrochimica Acta Part B: Atomic Spectroscopy*, vol. 65, no. 1, pp. 86–95, 2010.

[35] E. Mal, R. Junjuri, M. K. Gundawar et al., "Optimization of temporal window for application of calibration free-laser induced breakdown spectroscopy (CF-LIBS) on copper alloys in air employing a single line," *Journal of Analytical Atomic Spectrometry*, vol. 34, no. 2, pp. 319–330, 2019.

[36] G. Çelik, Ş. Ateş, and E. Erol, "Oscillator strengths and lifetimes for Cu I," *Canadian Journal of Physics*, vol. 93, no. 10, pp. 1015–1023, 2015.

[37] J. Hou, L. Zhang, W. Yin et al., "Development and performance evaluation of self-absorption-free laser-induced breakdown spectroscopy for directly capturing optically thin spectral line and realizing accurate chemical composition measurements," *Optics express*, vol. 25, no. 19, pp. 23024–23034, 2017.

[38] Q. Wang, A. Chen, W. Xu et al., "Effect of lens focusing distance on AlO molecular emission from femtosecond laser-induced aluminum plasma in air," *Optics & Laser Technology*, vol. 122, p. 105862, 2020.